\documentclass[12pt,preprint]{aastex}

\received{}
\accepted{}
\journalid{}{}
\articleid{}{}
\paperid{}
\cpright{AAS}{2002}
\ccc{}

\slugcomment{To be submitted to the {\it Astrophysical Journal}}

\shorttitle{The Molecular Line Opacity of MgH}
\shortauthors{P. F. Weck & al}

\begin{document}

\title{ The Molecular Line Opacity of MgH in Cool Stellar Atmospheres}

\author{P. F. Weck, A. Schweitzer, P. C. Stancil \& P. H. Hauschildt}
\affil{Department of Physics and Astronomy and Center for Simulational Physics, \\
The University of Georgia, Athens, GA 30602-2451}
\email{weck@-,andy@-,stancil@-,yeti@physast.uga.edu}

\author{K. Kirby}
\affil{Harvard-Smithsonian Center for Astrophysics\\
       60 Garden St., Cambridge, MA 02138}
\email{kirby@cfa.harvard.edu}


\begin{abstract}
A new, complete, theoretical rotational and vibrational line list for 
the $A~^2\Pi \leftarrow X~^2\Sigma^+$ electronic transition in
MgH is presented. The list includes transition energies and 
oscillator strengths for all possible allowed transitions and
was computed using the best available theoretical potential
energies and dipole transition moment function with the former
adjusted to account for experimental data. The $A\leftarrow X$ line
list, as well as new line lists for the 
$B'~^2\Sigma^+ \leftarrow X~^2\Sigma^+$ and the 
$X~^2\Sigma^+ \leftarrow X~^2\Sigma^+$ (pure rovibrational) transitions,
were included in comprehensive stellar atmosphere models for M, L, and
T dwarfs and solar-type stars. The resulting spectra, 
when compared to models lacking MgH,
show that MgH provides significant opacity in the visible between
4400 and 5600\AA~. Further, comparison of the spectra obtained with
the current line list to spectra obtained using the line list constructed
by \citet{kur93} show that the Kurucz list significantly overestimates
the opacity due to MgH particularly for the bands near 5150 and 4800\AA~
with the discrepancy increasing with decreasing effective temperature. 
\end{abstract}

\keywords{molecular data --- stars: atmospheres --- stars: late-type}

\section{Introduction}

The study of the spectra of cool stars requires detailed knowledge of 
molecular opacities. This includes important absorbers such as 
TiO, CO, and water vapor, which have bands that cover large wavelength 
ranges and are very important for the structure of the atmosphere due 
to their overall cooling or heating effects.

In addition, there are a number of molecules that have bands covering 
comparatively
small wavelength ranges (e.g., a few 10 or $100\,$\AA). Many of them are 
trace molecules that have only small effects on the overall physical conditions
inside the atmosphere but that are important for spectral classification and
for the determination of stellar parameters such as effective temperatures,
gravities and abundances. Unfortunately, important molecular data such
as energy levels, bound-free, and bound-bound cross-sections are only poorly
known or not known at all for a number of these trace molecules. We have therefore
started a project to update or provide for the first time molecular data
of astrophysical interest for important trace molecules and consider 
in this work MgH. These data will
be computed using state-of-the-art molecular physics codes and should improve
our ability to model and analyze cool stellar atmospheres considerably. 

It is important to assess the quality of the computed molecular data.
This is best done by comparing to experimental results; however, this is 
only possible for very few molecules of astrophysical interest. In addition,
in many cases the temperature range of astrophysical importance is higher than
what can be reached with current experimental setups. Therefore, indirect 
methods of testing and evaluating the molecular data are useful. In this paper,
we use the general-purpose stellar atmosphere code {\tt PHOENIX} to calculate
model atmospheres and synthetic spectra with and without the new molecular data.
The results of these calculations can then be used to assess the importance of 
particular molecular opacities on the structure of the atmosphere. The synthetic
spectra can be used to verify the correct strength of the computed bands when
compared to observational data. This procedure introduces uncertainties such as
the treatment of the equation of state (e.g., the molecular data used in it),
the treatment of lines and line profiles and the assumed parameters of the 
comparison star. However, differential analyses circumvent many of these 
problems and should allow a reasonable evaluation of the molecular bound-free
and free-free data. 

The electronic bands of magnesium hydride have been detected
over a wide range of stellar atmospheres including the photosphere
of the sun \citep{sot72}, sunspot umbrae \citep{wal99}, F-K giants in
the Milky Way halo and halos of other Local Group galaxies \citep{maj00},
and nearby L-dwarfs \citep{rei00}.
MgH lines can be used as indicators of surface gravity in
late-type stars \citep{bon93} and to determine magnesium isotope
abundances \citep{wal99,gay00}.

The spectrum of MgH has been extensively studied in the laboratory
for many decades \citep[and references therein]{bal76,bal78,ber85,wal99}
and has received some theoretical attention \citep[and references
therein]{sax78,kir79}. However, modern stellar atmosphere calculations
require extensive, and complete, molecular line lists as molecular
band absorption is the primary source of line-blanketing in cool
stellar atmospheres, particularly M dwarfs. For many molecules,
including MgH, the only source for complete line lists is 
the extensive compilations of \citet{kur93}. While these compilations 
are highly valuable
to stellar modelers, the methods necessary to compute 
100s of millions of lines require a number of approximations
which at times are severe. In this work, we apply fully quantum-mechanical
techniques to compute the complete line list for the $A~^2\Pi \leftarrow
X~^2\Sigma^+$ transition of MgH. The parameters of the calculation are
adjusted to force agreement with available experiments. However,
our goal is to reproduce the global MgH opacity, hence we cannot claim
spectroscopic accuracy for a particular line. The $A\leftarrow X$ line
list constructed in this work is combined with line lists for
the $B'~^2\Sigma^+ \leftarrow X~^2\Sigma^+$ and the
$X~^2\Sigma^+ \leftarrow X~^2\Sigma^+$ transitions computed by
\citet{sko02} and tested in a range of stellar atmosphere models.
An overview of the theory of molecular rotational lines is presented
in section 2 with the results of the line list calculations
and stellar models are given in section 3. We present our
conclusions in section 4.

\section{Molecular theory}

\subsection{Potential curves and dipole transition moments} \label{pot}

For the purpose of the present calculations, accurate {\em ab initio} 
potential-energy surfaces given by \cite{sax78} have been used for both the 
$A~^2\Pi$ and $X~^2\Sigma^+$ electronic states of MgH. Their configuration 
interaction treatment includes singly- to 
triply-excited configurations (SDTCI) in a large Slater-type basis set, leading 
to good agreement of the derived spectroscopic constants with experimental data. 

In order to bring the calculated potential curves into better agreement 
with experiment, shifts of 
$+8.740\times 10^{-4}~\mbox{a.u.}$\footnote{Atomic units are used throughout this 
section unless otherwise stated.} 
and $-3.3665\times 10^{-3}~\mbox{a.u.}$ 
have been applied in the 
present work to the \cite{sax78} $X~^2\Sigma^+$ and $A~^2\Pi$ energies, 
respectively. Details about this procedure are given in a separate publication for the 
$B'\leftarrow X$ electronic transition in MgH \citep{sko02}. 
 In this way, the dissociation energies $D_0^0$ coincide with 
the experimental values given by \cite{bal78} for 
the ground state and by \cite{bal76} for the $A~^2\Pi$ state. The
dissociation energies adopted here are 10,243.26 and and 12,903.71 cm$^{-1}$
for the $X~^2\Sigma^+$ and $A~^2\Pi$ of $^{24}$MgH, respectively. 
The relative energies between the two potential curves were further
shifted to match the energy difference corresponding to the 
pure $(0,0)$ vibronic transition. The value 
of $T_0=19278.13~\mbox{cm}^{-1}$, which corresponds to the average of 
the Q$_1$(0) and Q$_2$(1) transition energies measured by \cite{bal76}, 
was adopted. 
We note that this value for $T_0$ does not agree with the more
recent measurements of Bernath, Black, \& Brault (1985) who
find $T_0=19284.65$ cm$^{-1}$, though it is unclear how their
value was arrived at.  
In fact, their assumption of a Hund's coupling case (a) 
is in obvious contradiction with the small and regular 
$\Lambda$-doubling parameters,  
$p_0=0.0258~\mbox{cm}^{-1}$ and $q_0=0.00178~\mbox{cm}^{-1}$.
Such a case for a $^2\Pi\leftarrow ^2\Sigma^+$ transition
is more typically characterized by Hund's case (b) 
\citep*[Figures 122 and 124]{her50}. Further, \citep{ber85}
suggest that a number of low-$J$ lines, presumably $P_1$(1), $R_2$(1),
$Q_2$(0), $Q_2$(1), $Q_2$(2), and $P_2$(2), measured by
\citet{bal76} do not exist apparently assuming Hund's case (a).
However, we find that under Hund's case (b) all but $P_1$(1) and
$Q_2$(0) do in fact exist. We therefore adopted the value
of $T_0$ deduced from \citet{bal76}.       

Out of the range of internuclear separations $R=2.2~\mbox{a}_o$ to 
$9.5~\mbox{a}_o$ 
considered by \cite{sax78}, the potential curves have been extrapolated 
in two different ways. On the one hand, for internuclear distances 
$R> 9.5~\mbox{a}_o$, a smooth fit to the {\it ab initio} potentials has been 
performed using the average long-range interaction potential    
\begin{equation}
V_{\rm L} (R) = - {C_{6}\over{R^{6}}}- {C_{8}\over{R^{8}}}
- {C_{10}\over{R^{10}}},
\end{equation}
where $C_6$, $C_8$ and $C_{10}$ are the usual van der Waals coefficients 
corresponding to the dipole-dipole, dipole-quadrupole, and the sum of 
dipole-quadrupole and dipole-octupole interactions, respectively. 
For the $X~^2\Sigma^+$ ground state, the coefficients described in \citet{sko02}
were adopted as summarized in Table \ref{tbl1}. 
Since to our knowledge, no data have been reported 
for the van der Waals constants of the $A~^2\Pi$ state of MgH, they  
were estimated using a technique based on the London formula 
as described for the $B'~^{2}\Sigma^+$ state by 
\cite{sko02}. On the other hand, for internuclear distances 
$R < 2.2~\mbox{a}_o$, the potential curves of both the $X~^2\Sigma^+$ and the 
$A~^2\Pi$ electronic states have been fitted to the short-range 
interaction exponential form $A\exp(-BR) + C$.

In a similar way, the dipole transition moment for 
$A~^2\Pi\leftarrow X~^2\Sigma^+$ 
and the dipole moment of the $X$ state calculated by \cite{sax78} have been 
used over the range $R=2.2~\mbox{a}_o$ to $9.5~\mbox{a}_o$, and extrapolated 
by an exponential 
fit for both the short- and long-range interactions. The dipole transition  
moment and the dipole moment were both smoothly forced to zero at the 
united-atom and separated-atom limits.

\subsection{Rotational and band oscillator strengths} \label{os}

Throughout the present study, we have adopted the now well established point 
of view expressed by \cite{whi74} with respect to the preferred way in which 
dipole matrix elements of a rotational transition should be separated into 
rotational and electronic-vibrational parts. 

For the $A~^2\Pi\leftarrow X~^2\Sigma^+$ absorption band system of 
the present study, the rotational oscillator 
strength can be expressed as given by \cite{lar83}, 
\begin{equation}\label{fbr}
f^{ab}_{v'J',v''J''} = \frac{2}{3} \Delta E_{v'J',v''J''}\frac{S_{J'}(J'')}{2J''+1}
\vert D^{AX}_{v'J',v''J''}\vert ^2,
\end{equation}
where 
$D^{AX}_{v'J',v''J''}=<\chi_{v'J'}\vert D^{AX}(R)\vert\chi_{v''J''}>$ 
is the rovibrational matrix element of the electric dipole 
transition moment $D^{AX}(R)$ responsible 
for absorption from the $X~^2\Sigma^+$ 
into the $A~^2\Pi$ electronic state and $\chi_{vJ}$ are the
rovibrational wave functions.
The H\"{o}nl-London factors $S_{J'}(J'')$ are defined according 
to \cite{whi74} as 
\begin{eqnarray}
S_{J'}(J'')&=&
\left\{
\begin{array}{ll}
(J''-1)/2,&~J'=J''-1~(\mbox{P-branch})\\
(2J''+1)/2,&~J'=J''~~~~~~(\mbox{Q-branch})\\
(J''+2)/2,&~J'=J''+1~(\mbox{R-branch}).
\end{array} \right .   
\end{eqnarray}

For the sake of comparison with the band oscillator strength values calculated 
by \cite{kir79}, we used the following relation between band and rotational 
oscillator strengths, as defined in Eq.(\ref{fbr}), 
\begin{eqnarray}\label{axf} 
f^{ab}_{v'v''} &=& \frac{1}{g^{ab}_{J',J''}S_{J'}(J'')}~f^{ab}_{v'J',v''J''}
\nonumber \\
 &=& \frac{(2-\delta_{0,\Lambda''+\Lambda'})(2J''+1)}
{(2-\delta_{0,\Lambda''})S_{J'}(J'')}~f^{ab}_{v'J',v''J''},
\end{eqnarray}
where $g^{ab}_{J',J''}$ is a degeneracy factor arising from 
spin-splitting and $\Lambda$-doubling in both final and initial electronic 
states.  


\section{Results and discussion}

\subsection{The $A-X$ rovibrational line list}

The energy levels together with the rovibrational wavefunctions 
$\chi_{v'J'}(R)$ and $\chi_{v''J''}(R)$ of the final and initial 
electronic state have been calculated by solving the radial nuclear 
equation by standard Numerov techniques \citep{coo61}. 
These calculations have been 
performed using a step of $1\times10^{-3}~\mbox{a.u.}$ 
for the integration over internuclear distances from $R=0.5~\mbox{a.u.}$ 
to $200~\mbox{a.u.}$ For $^{24}$MgH, the reduced mass 0.9671852~u\footnote{In 
atomic weight units, Aston's scale} =
1763.064~a.u. \citep{hub79} was adopted.  

In Table \ref{tbl2}, the vibrational binding energies of the $A~^2\Pi$ 
state calculated for the present study are given together with the 
corresponding values of $\Delta G(v'+1/2)=G(v'+1)-G(v')$. 
The latter are found to be in excellent agreement with the previous 
calculations of \citet{sax78} up to $v'=4$. For higher vibrational levels, a 
maximum difference of 125 cm$^{-1}$ occurs for $v'=8$. Moreover, our fitting 
procedure to the long-range interaction potential yields an additional 
vibrational level near threshold to give $v'_{\rm max}=13$. The maximum
$J'$ for each $v'$ is also shown in Table~\ref{tbl2} giving a total of
435 rovibrational states.\footnote{A discussion of the number and accuracy
of the $X~^2\Sigma^+$ rovibrational states is given in \citet{sko02}.} 
Comparison to the limited experimental
data shows the calculated $\Delta G$'s to be smaller by up to 45 cm$^{-1}$
which suggests that the \citet{sax78} potential is somewhat broader than 
reality.      

The band oscillator strength values for the band system 
$A~^2\Pi\leftarrow X~^2\Sigma^+$ are given in Table \ref{tbl3} for 
the transitions between the vibrational states $v''=0-3$ and $v'=0-6$. 
The current calculations satisfactorily reproduce the theoretical 
results of \citet{kir79}. 

\begin{figure}
\plotone{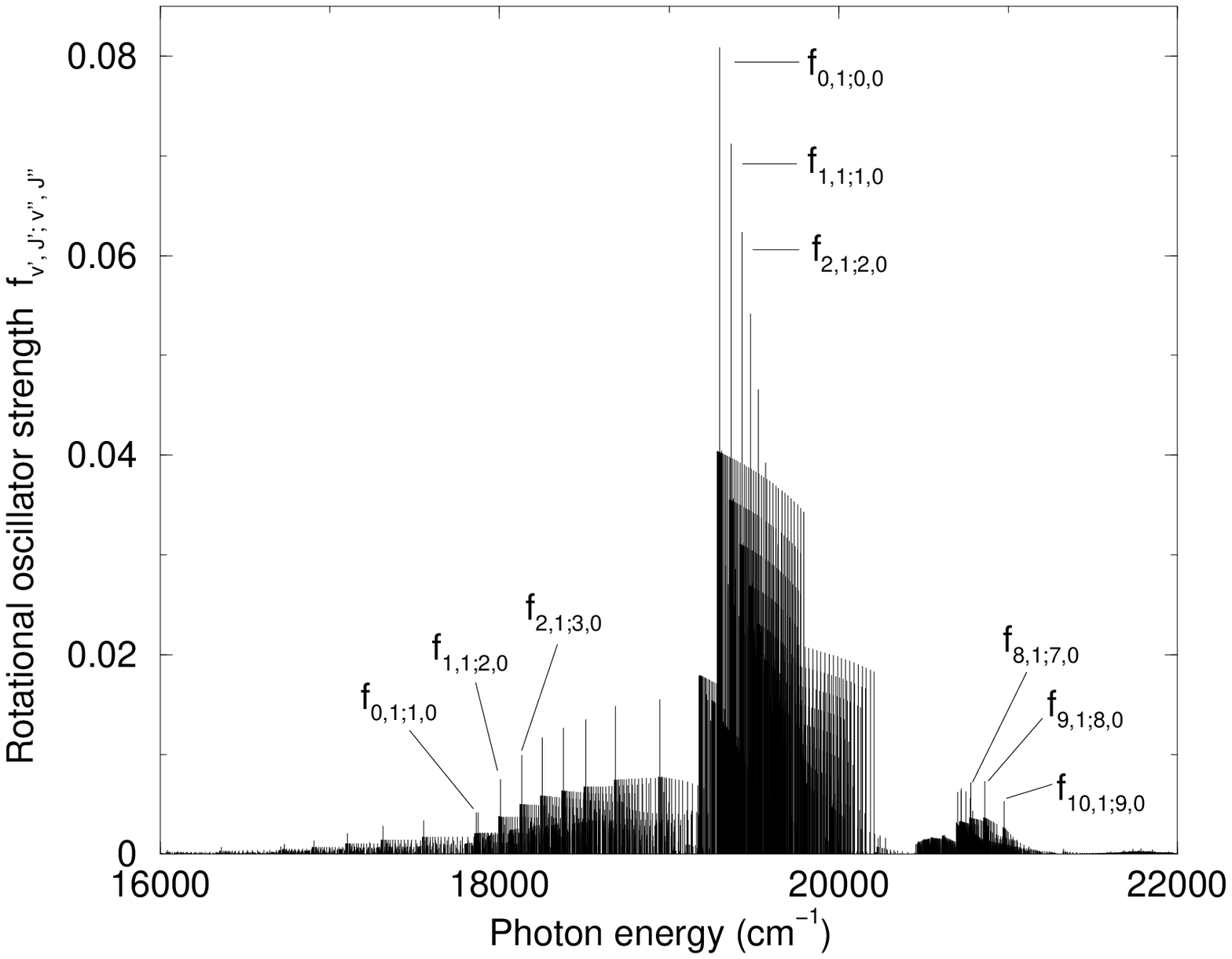}
\caption{Rotational oscillator strengths $f_{v'J',v''J''}$ as a function of 
 wavelength of the absorbed photon energy.\label{fig1}}
\end{figure}

Figure \ref{fig1} shows the line absorption rotational oscillator 
strengths\footnote{The complete list
of MgH oscillator strength data
is available online at the UGA Molecular Opacity Project database website
\url{http://www.physast.uga.edu/ugamop/}} for the 
$A~^2\Pi\leftarrow X~^2\Sigma^+$ transition as a function of the energy of the 
absorbed photon. It is worth noting the presence of a series of peaks 
which correspond to the $J'=1\leftarrow J''=0$ lines of the R-branch 
[$R(0)$] whose 
intensities are larger by a factor two than all other lines in the
same wavelength region. A detailed examination of the wavelength 
distribution of these peaks reveals that 
they are to be classified into three categories, according to whether they 
are $\Delta v=v'-v''=+1,~0~\mbox{or} -1$ transitions as shown in 
Figure \ref{fig1}. Apart from these peaks, the R-branch contribution dominates 
over the wavelength ranges $19,800-20,300~\mbox{cm}^{-1}$ and 
$21,050-21,300~\mbox{cm}^{-1}$. Other lines 
observed in Figure \ref{fig1} are to be 
assigned mainly to the Q-branch transitions, which are about two times more 
intense than the P-branch over the whole wavelength range considered. However, 
the P-branch appears to play a major role for the energy range
$19,170-19,280~\mbox{cm}^{-1}$ and $20,450-20,690~\mbox{cm}^{-1}$.         

\begin{figure}
\plotone{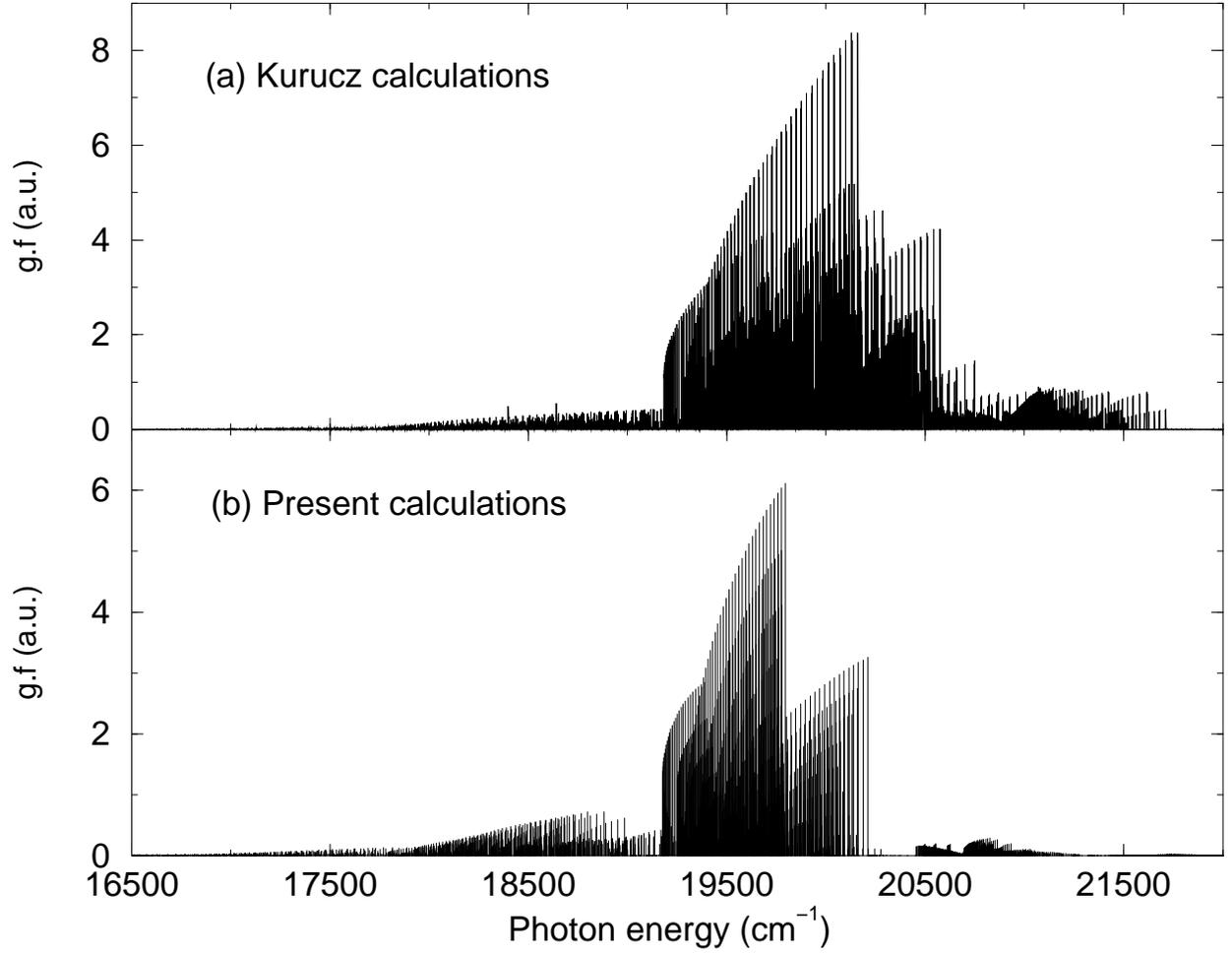}
\caption{$g_{J',J''}.f_{v'J',v''J''}$ values as a function of the absorbed 
 photon energy. (a) calculations of \cite{kur93}; (b) present 
 calculations.\label{fig2}}
\end{figure}

The product $g_{J',J''}.f_{v'J',v''J''}$, the so-called $gf$-value, 
is plotted in Figure \ref{fig2} as a
function of the absorbed photon energy were comparison is made to
the line list of \citet{kur93}. 
To the best of our knowledge, the \cite{kur93} calculations 
were performed with a model 
rotational Hamiltonian using spectroscopic constants
\citep{kur93b}. Thus, the 
differences arising between both sets of results can be mainly explained by 
the fact that the model Hamiltonian method, though satisfactory
for low-lying rovibrational 
levels, may lose accuracy as $v$ and/or $J$ 
increase. Further, it seems that the same $J_{\rm max}$ was used for all
vibrational levels (Kurucz, private communication, 2002) which is clearly
not the case as shown in Table~\ref{tbl2}. This appears to explain why the
band-series in the \citet{kur93} calculations extend to larger photon energies.

\subsection{Atmosphere models}

The  models  used for this  work were calculated as described in
\cite{LimDust}.
These models  and their comparisons to earlier versions  are the
subject of a  separate publication \cite[]{LimDust} and we thus do not repeat
the detailed description of the models here.
However, we will briefly summarize the major physical properties.
The models are based on the Ames H$_2$O  and TiO line lists by
\cite{ames-water-new} and \cite{ames-tio}
and also include as a new addition the line lists for FeH by \cite{FeHberk2}
and for VO and CrH by R. Freedman (NASA-Ames, private communication).
The models account for equilibrium formation of dust and condensates
and include grain opacities for 40 species.
In the following, the models
will be referred to as ``AMES-dusty'' for models in which the dust particles
stay in the layers in which they have formed and ``AMES-cond'' for models in
which the dust particles have sunk below the atmosphere from the layers in
which they originally formed.
We  stress  that large
uncertainties persist  in the water opacities for parts of the  temperature range of this
work \citep{2000ApJ...539..366A}. However, almost all MgH bands are
in the optical and are thus affected minimally by the quality of
the water opacities.

In addition to the opacity sources listed above and in \citet[and references
therein]{LimDust} we added the new 
$A$-$X$ bound-bound radiative transition data from this work and the $B'$-$X$ 
and $X$-$X$ data from \citet{sko02}
for all isotopes $^{24}$MgH, $^{25}$MgH, and $^{26}$MgH
to our opacity database.
In order to assess the effects of the new MgH data,
we compare spectra calculated with these opacity sources to spectra calculated
with the MgH line data provided by Kurucz in his list of
spectral lines of  diatomic molecules \citep{kur93}.
In addition,
we have computed a number of models with and without MgH data.
The original AMES grid was calculated for effective temperatures of M, L, and
T~dwarfs. The hotter models in this work are based on the same physics
as the AMES grid and merely differ in the effective temperature.
The models used in the following discussion were all
iterated to convergence for the parameters indicated.
The high resolution spectra which have the individual opacity
sources  selected are calculated on top of the models.
The MgH bands are too localized in a region with
little flux, or too weak to influence the temperature
structure of the atmosphere.

\begin{figure}
\plotone{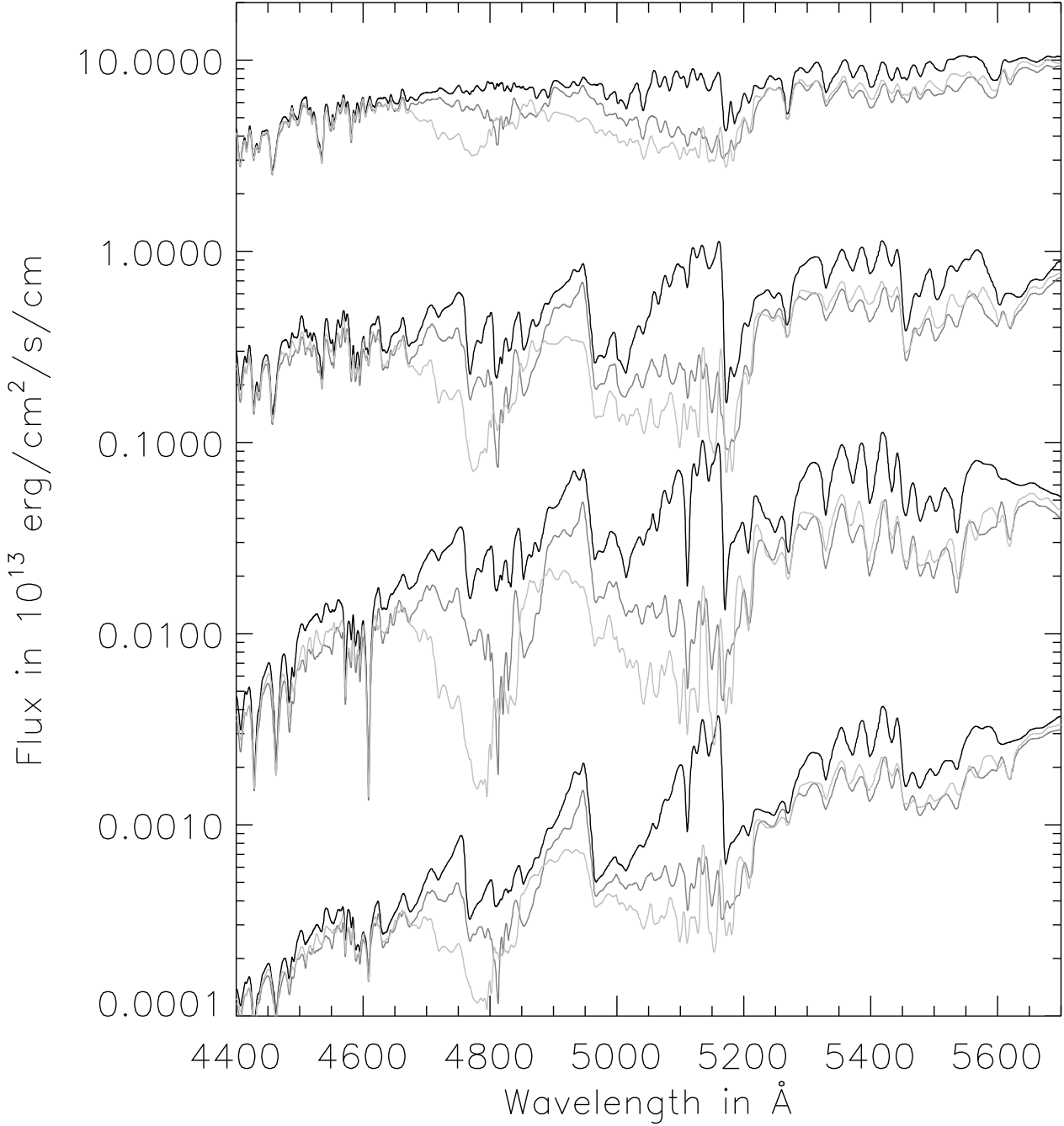}
\caption{Spectra of the strongest MgH bands for models with T$_{\rm eff}$=4000~K,
T$_{\rm eff}$=3000~K, T$_{\rm eff}$=2000~K (AMES-Cond model) and T$_{\rm eff}$=2000~K
(AMES-Dusty model) (from top to bottom). The 2000~K AMES-Cond models have an
artificial offset of +1.0~dex for better legibility.  All models have log(g)=5.0
and are AMES-Cond unless noted otherwise.  The black spectra are from models
without MgH data, the dark grey spectra are from models with the new MgH data
and the light grey spectra are from  models using the \citet{kur93} 
MgH data.  The
resolution of the calculated spectra are much higher. For illustrative purposes
the spectra have been reduced to a resolution of R=80 at 5000~\AA.
\label{figspec}}
\end{figure}

\begin{figure}
\plotone{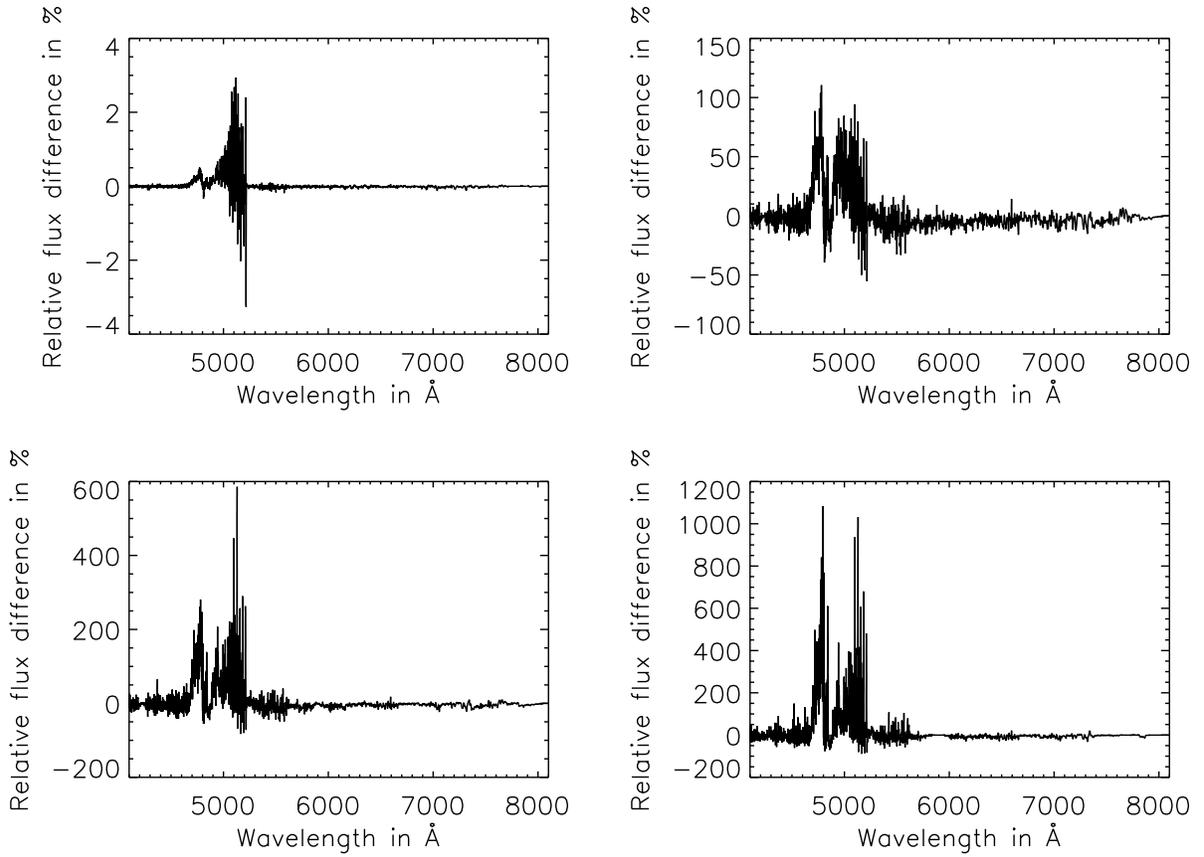}
\caption{Relative flux differences between models when changing from the 
\citet{kur93} MgH data to the new MgH data. 
The models are from top left to bottom right
5800~K, 4000~K, 3000~K and 2000~K. All models are AMES-Cond and have
log(g)=5.0, except the 5800~K model has log(g)=4.5. The resolution of all
spectra have been reduced to R=500 before calculating the ratio.
\label{figdiff}}
\end{figure}

In Figure \ref{figspec} we show comparisons between
spectra from models using  no MgH data, spectra from models using
the new MgH line list and spectra from
models using the \citet{kur93} MgH line list in the spectral region where the
MgH bands are most prominent. The comparisons have been done
at effective temperatures of 2000~K, 3000~K, and 4000~K to sample the
temperature range in which MgH is visible in the spectrum.
For convenience, we only did the comparison for models with log(g)=5.0.
As can be seen, the \citet{kur93} line lists overestimates the opacity due
to the inclusion of non-existent levels with high $J$-values and the
use of a model Hamiltonian approach.
The differences between the spectra with the new and \citet{kur93}
MgH line lists are
very similar for the AMES-Cond and AMES-Dusty models
although the overall flux level and the overall flux
shape are different for AMES-Cond and AMES-Dusty in the
optical (Note that reversing the 1.0~dex offset applied in the plot
is not sufficient to match the two 2000~K models).
Another view is presented in figure \ref{figdiff}
which shows the relative difference in the spectra.
As can be seen,
there are significant differences spreading among both the
$A-X$ and the $B'-X$ transitions.

Since the MgH data presented by \citep{kur93} was intended for solar-type
stars we also compared the resulting spectra for an effective
temperature of 5800~K and log(g)=4.5, relevant parameters for solar-type
stars. As can be seen as part of Figure \ref{figdiff}, 
the differences are negligible.
Although the effect of the non-existent levels with high $J$-values should be  
more important for hotter
temperatures, the decrease in the abundance of MgH at higher temperatures 
makes the effect unobservable.
Finally, we want to note that for the 2000~K AMES-Cond 
model the X-X band adds a small amount of opacity at around 
4\micron~ and 10\micron.

\section{Conclusion}

Using a combination of theoretical and experimental data
on the potential energies and dipole transition moment of
MgH, a comprehensive theoretical vibrational-rotational line list for the 
$A~^2\Pi \leftarrow X~^2\Sigma^+$ transition was constructed.
When using the new $A\leftarrow X$ line data and the new
$B'~^2\Sigma^+ \leftarrow X~^2\Sigma^+$ and the
$X~^2\Sigma^+ \leftarrow X~^2\Sigma^+$ line data of \cite{sko02}  
in synthetic spectrum calculations,
we find significant differences in the opacity when comparing the spectra
to calculations using the existing data of \citet{kur93}.
The differences are largest for effective temperatures pertaining to 
L and M type stars and can easily be seen in low resolution work.   
For hotter stars, of K and G type,
the differences are less pronounced and high resolution
spectra are required to notice the improvements for the hottest stars.

\acknowledgments

This work was supported in part by NSF grants AST-9720704 and AST-0086246,
  NASA
grants NAG5-8425, NAG5-9222, and NAG5-10551 as well as NASA/JPL 
grant 961582 to the University
of Georgia. This work also was supported in
part by the P\^ole Scientifique de Mod\'elisation Num\'erique at ENS-Lyon.
Some of the calculations presented in this paper were performed on the IBM 
SP2 and the SGI Origin
of the UGA EITS, on the IBM SP ``Blue Horizon'' of the San Diego 
Supercomputer
Center (SDSC), with support from the National Science Foundation, and on 
the
IBM SP of the NERSC with support from the DoE.  We thank all these 
institutions
for a generous allocation of computer time.
We thank Stephen Skory for assistance during the early part of this work.
P.F.W. and K.K. are grateful to B. Kurucz for useful discussions. 



\clearpage

\clearpage

\clearpage

\clearpage

\clearpage

\begin{deluxetable}{llccrrrl}
\tablecaption{Asymptotic Separated-Atom and United-Atom Limits\label{tbl1}}
\tablecolumns{8}
\tablehead{
\multicolumn{1}{c}{} & \multicolumn{6}{c}{Separated-Atom} & \multicolumn{1}{c}{} \\
\cline{2-7}
\multicolumn{1}{l}{Molecular} &\multicolumn{1}{c}{} & \multicolumn{2}{c}{Energy (eV)} & \multicolumn{3}{c}{} &
\multicolumn{1}{c}{United}\\
\cline{3-4}
\multicolumn{1}{c}{State} & \multicolumn{1}{c}{Atomic States} &
\multicolumn{1}{c}{Theory} & \multicolumn{1}{c}{Expt} & \multicolumn{1}{c}{C$_6$} &
\multicolumn{1}{c}{C$_8$} & \multicolumn{1}{c}{C$_{10}$} &
\multicolumn{1}{c}{Atom (Al)} }
\startdata
$X$ $^2\Sigma^+$ & Mg($3s^2~^1 S$)+H($1s~^2 S$) & 0.0 & 0.0 & 57.8\tablenotemark{a} & 2490\tablenotemark{b} & 115000\tablenotemark{b} & $3s^23p~^2 P^0$ \\
$A$ $^2\Pi$ & Mg($3s3p~^3 P^0$)+H($1s~^2 S$) & 2.596 & 2.714\tablenotemark{c} & 56.9\tablenotemark{d} & 2451\tablenotemark{d} & 113205\tablenotemark{d} & $3s^23p~^2 P^0$ \\
$B'$ $^2\Sigma^+$ & Mg($3s3p~^3 P^0$)+H($1s~^2 S$) & 2.593 & 2.714\tablenotemark{c} & 56.9\tablenotemark{d} & 2451\tablenotemark{d} & 113205\tablenotemark{d} & $3s^24s~^2 S$ \\
\enddata 
\tablenotetext{a}{A. Derevianko (private communication, 2001).}
\tablenotetext{b}{From \citet{std85}.}
\tablenotetext{c}{Weighted average of the $^3 P_0$, $^3 P_1$, and $^3 P_2$ term energies from \citet{nis99}.}
\tablenotetext{d}{Estimate, \citet{sko02}.}
\end{deluxetable}


\clearpage

\begin{deluxetable}{rccrrr} 
\tablecolumns{6} 
\tablewidth{0pc} 
\tablecaption{Vibrational binding energies\tablenotemark{a}~ and 
 $\Delta G(v'+1/2)$ in $\mbox{cm}^{-1}$ for the $A~^2\Pi$ state \label{tbl2}} 
\tablehead{ 
\colhead{} &\colhead{} &\colhead{} & \multicolumn{3}{c}{$\Delta G(v'+1/2)$} \\ 
\cline{4-6} \\ 
\colhead{$v'$} & \colhead{$J'_{max}$} & \colhead{B. E.\tablenotemark{b}} 
& \colhead{Theory\tablenotemark{b}} & \colhead{Saxon\tablenotemark{c}} & 
\colhead{Expt.\tablenotemark{d}}}    
\startdata 
0 & 49 & 12903.7 & 1494.3 & 1493.9 & 1533.9\\
1 & 47 & 11409.4 & 1423.2 & 1423.2 & 1466.1\\
2 & 44 & 9986.2 & 1349.0 & 1349.5 & 1394.4\\
3 & 42 & 8637.2 & 1273.2 & 1273.0 & \nodata\\
4 & 39 & 7364.0 & 1196.4 & 1191.8 & \nodata\\ 
5 & 36 & 6167.5 & 1132.0 & 1103.9 & \nodata\\
6 & 33 & 5035.5 & 1082.9 & 1007.4 & \nodata\\
7 & 30 & 3952.6 & 1013.1 & 900.2 & \nodata\\
8 & 27 & 2939.5 & 905.7 & 780.2 & \nodata\\
9 & 23 & 2033.8 & 744.7 & 645.7 & \nodata\\
10 & 19 & 1289.1 & 573.3 & 495.0 & \nodata\\ 
11 & 15 & 715.8 & 414.1 & 330.9 & \nodata\\
12 & 11 & 301.8 & 170.1 & \nodata & \nodata\\
13 & 6 & 131.7 & \nodata & \nodata & \nodata\\
\enddata 

\tablenotetext{a}{Binding energies are given for rotationless 
vibrational levels} 
\tablenotetext{b}{This work}
\tablenotetext{c}{\cite{sax78}}
\tablenotetext{d}{\cite{bal76}}
\end{deluxetable} 


\clearpage

\begin{deluxetable}{rrrrrrrrr} 
\tablecolumns{9} 
\tablewidth{0pc} 
\tablecaption{Band oscillator strengths\tablenotemark{\star} ~and 
transitions energies\tablenotemark{\dagger} ~for the $A~^2\Pi\leftarrow X~^2\Sigma^+$ 
Band System \label{tbl3}} 
\tablehead{ 
\colhead{}  &\multicolumn{2}{l}{$(X^2\Sigma^+) ~v''=0$}
 & \multicolumn{2}{c}{$v''=1$}
 & \multicolumn{2}{c}{$v''=2$}
 & \multicolumn{2}{c}{$v''=3$}\\
\colhead{$v'~(A^2\Pi)$} &\colhead{$E_{v'0}$} &\colhead{$f_{v'0}$} & 
\colhead{$E_{v'1}$} &\colhead{$f_{v'1}$}  &
\colhead{$E_{v'2}$} &\colhead{$f_{v'2}$}  &
\colhead{$E_{v'3}$} &\colhead{$f_{v'3}$} 
}    
\startdata 
0............ &19290 &1.616(-1)& 17867 &8.30(-3) &16506  &5.2(-4) &15209 &4.1(-5)\\
 &19292 &1.61(-1) &17860 &8.61(-3) &16492 &5.3(-4) &15190 &4.1(-5)\\

1............ &20784 &4.88(-3) & 19361 &1.424(-1) &18000 &1.49(-2) &16702
 &1.48(-3)\\
 &20826 &5.14(-3) &19394 &1.41(-1) &18025 &1.59(-2) &16724 &1.52(-3)\\
 
2............ &22207 &1.0(-5) & 20784 &8.62(-3) &19422 &1.246(-1) &18125 &1.99(-2)\\

 &22288 &3.9(-6) &20856 &9.38(-3) &19490 &1.24(-1) &18185 &1.99(-2)\\
 
3............ &23555 &6.9(-7) & 22132 &5.9(-5) &20770 &1.11(-2) &19473 &1.082(-1)\\
 &23682 &4(-7) &22250 &4.4(-5) &20884 &1.14(-2) &19579 &1.09(-1)\\
 
4............ &24828 &1.4(-8) & 23405 &2.2(-6) &22043 &2.0(-4) &20747 &1.25(-2)\\
 &25002 & $<1.0(-7)$ &23570 & 7.0(-6) &22204 &2.4(-4) &20899 &1.27(-2)\\
 
5............ &26024 &1.3(-9) & 24601 &1.3(-7) &23240 &2.8(-6) &21943 &5.0(-4)\\
 &26243 & $<1.0(-7)$ &24811 & $<1.0(-7)$ &23445 &5.0(-6) &22140 &5.2(-4)\\
 
6............ &27156 &1.2(-9) & 25733 &8.7(-9) &24371 &6.5(-7) &23074 &1.2(-6)\\
 &27398 & $<1.0(-7)$ &25966 & 4.0(-7) &24600 &2.0(-6) &23295 &1.0(-6)\\
\enddata 

\tablenotetext{\star}{Band oscillator strengths are given in a.u. 
 and are calculated for rotational quantum numbers $J''=0$ and $J'=1$. 
 Our results are listed on the first line, with the previous calculations of 
  \cite{kir79} below. Notation: $x(-n)\equiv x\times 10^{-n}$}
\tablenotetext{\dagger}{Energies $E_{v'v''}$ of the absorbed photon are 
 given in $\mbox{cm}^{-1}$.
 Our values are listed on the first line, with the experimental 
 transition energies given by \cite{bal76} below (generally as listed
 in \citet{kir79}), for $J''=0$ and $J'=1$.} 

\end{deluxetable}





\begin{thebibliography}{}

\bibitem[\protect\citeauthoryear{Allard, Hauschildt, Alexander, Tamanai \&
  Schweitzer}{Allard et~al.}{2001}]{LimDust}
    Allard F.,  Hauschildt P.~H.,  Alexander D.~R.,  Tamanai A.,    
    Schweitzer A., 2001, ApJ, 556, 357
\bibitem[\protect\citeauthoryear{{Allard}, {Hauschildt} \&
  {Schweitzer}}{{Allard} et~al.}{2000}]{2000ApJ...539..366A}
  {Allard} F.,  {Hauschildt} P.~H.,    {Schweitzer} A.,  2000, ApJ, 539, 366
\bibitem[Balfour \& Cartwright(1976)]{bal76} Balfour, W. J.,
   \& Cartwright, H. M.  1976, \aaps, 26, 389 
\bibitem[Balfour \& Lindgren(1978)]{bal78} Balfour, W. J.,
   \& Lindgren, B.  1978, Can. J. Phys., 56, 767
\bibitem[Bernath et al.(1985)]{ber85} Bernath, P. F.,
   Black, J. H., \& Brault, J. W.  1985, \apj, 298, 375
\bibitem[Bonnell \& Bell(1993)]{bon93} Bonnell, J. T. \& Bell, R. A. 1993,
   \mnras, 264, 334
\bibitem[Cooley(1961)]{coo61} 
   Cooley, J. W.  1961, Math. Computation, 15, 363
\bibitem[Gay \& Lambert(2000)]{gay00} Gay, P. L.  \& Lambert, D. L. 2000,
   \apj, 533, 260
\bibitem[Herzberg(1950)]{her50} 
   Herzberg, G.  1950, Molecular Spectra and Molecular Structure, 
    Vol. I, Spectra of Diatomic Molecules 
   (Princeton: D. Van Nostrand)
\bibitem[Huber \& Herzberg(1979)]{hub79} Huber, K. P. \&
   Herzberg, G.  1979, Molecular Spectra and Molecular Structure,
    Vol.  IV, Constants of Diatomic Molecules (New York: Van Nostrand Reinhold)
\bibitem[Kirby et al.(1979)]{kir79} Kirby, K.,
   Saxon, R. P., \& Liu, B.  1979, \apj, 231, 637
\bibitem[Kurucz(1993a)]{kur93} Kurucz, R. L.  1993a, 
   CD-ROM No.15 Diatomic molecular data for opacity calculations 
   (Harvard-Smithsonian Center for Astrophysics)   
\bibitem[Kurucz(1993b)]{kur93b} Kurucz, R. L.  1993b,
     Molecules in the Stellar Environment, ed. U. G. J{\o}rgensen
     (Berlin: Springer-Verlag), p. 282  
\bibitem[Larsson(1983)]{lar83} 
   Larsson, M.  1983, \aap, 128, 291
\bibitem[Majewski et al.(2000)]{maj00} Majewski, S. R., Ostheimer, J. C.,
   Kunkel, W. E., \& Patterson, R. J. 2000, \aj, 120, 2550
\bibitem[NIST Atomic Spectra Database(1999)]{nis99} NIST Atomic 
      Spectra Database 1999,
      http://aeldata.phy.nist.gov/cgi-bin/AtData/main\_asd
\bibitem[\protect\citeauthoryear{Partridge \& Schwenke}{Partridge \&
  Schwenke}{1997}]{ames-water-new}
  Partridge H.,  Schwenke D.~W.,  1997, J. Chem. Phys., 106, 4618
\bibitem[\protect\citeauthoryear{Phillips \& Davis}{Phillips \&
  Davis}{1993}]{FeHberk2}
  Phillips J.~G.,  Davis S.~P.,  1993, ApJ, 409, 860
\bibitem[Reid et al.(2000)]{rei00} Reid, I. N., Kirkpatrick, J. D.,
 Gizis, J. E., Dahn, C. C.,
 Monet, D. G., Williams, R. J.,
 Liebert, J., \& Burgasser, A. J. 2000, \aj, 119, 369
\bibitem[Saxon et al.(1978)]{sax78} Saxon, R. P.,
   Kirby, K., \& Liu, B.  1978, \jcp, 12, 5301 
\bibitem[\protect\citeauthoryear{Schwenke}{Schwenke}{1998}]{ames-tio}
  Schwenke D.~W.,  1998, Chemistry and Physics of Molecules and Grains in Space.
  Faraday Discussion, 109, 321
\bibitem[Skory et al.(2002)]{sko02} Skory, S., Stancil, P. C., 
   Weck, P. F., \& Kirby, K. 2002, \apjs, in preparation 
\bibitem[Sotirovski(1972)]{sot72} Sotirovski, S. 1972, \aaps, 6, 85
\bibitem[Standard \& Certain(1985)]{std85} Standard, J. M., \& 
   Certain, P. R.  1985, J. Chem. Phys., 83, 3002 
\bibitem[Wallace et al.(1999)]{wal99} Wallace, L., Hinkle, K.,
   Li, G., \& Bernath, P. 1999, \apj, 524, 454
\bibitem[Whiting \& Nicholls(1974)]{whi74} Whiting, E. E., \& 
   Nicholls, R. W.  1974, \apjs, 27, 1    
   

\end{thebibliography}
\end{document}